# Second-harmonic generation induced by electric currents in GaAs

Brian A. Ruzicka,[1] Lalani K. Werake,[1] Guowei Xu,[1]
Jacob B. Khurgin,[2] E. Ya. Sherman,[3] Judy Z. Wu[1], and Hui Zhao[1]*

[1]*Department of Physics and Astronomy, The University of Kansas, Lawrence, Kansas 66045, USA*
[2]*Department of Electrical and Computer Engineering,
Johns Hopkins University, Baltimore, Maryland 21218, USA*
[3]*Departamento de Quimica Fisica, Universidad del Pais Vasco-Euskal Herriko Unibertsitatea,
48080 Bilbao, Spain and IKERBASQUE, Basque Foundation for Science, 48011 Bilbao, Spain*
(Dated: December 21, 2011)

We demonstrate a new, nonlinear optical effect of electric currents. First, a steady current is generated by applying a voltage on a doped GaAs crystal. We demonstrate that this current induces second-harmonic generation of a probe laser pulse. Second, we optically inject a transient current in an undoped GaAs crystal by using a pair of ultrafast laser pulses, and demonstrate that it induces the same second-harmonic generation. In both cases, the induced second-order nonlinear susceptibility is proportional to the current density. This effect can be used for nondestructive, noninvasive, and ultrafast imaging of currents. These advantages are illustrated by the real-time observations of a coherent plasma oscillation and spatial resolution of current distribution in a device. This new effect also provides a mechanism for electrical control of the optical response of materials.

Interaction between matter and electromagnetic field is usually described in the coordinate gauge with the electric field strength and the polarization of the matter. One important example of this interaction in the nonlinear regime is the well-known electric-field-induced second-harmonic generation (SHG) effect,[1] in which a low frequency electric field breaks the inversion symmetry and, in the presence of an optical field with frequency $\omega$, engenders output of a second harmonic (SH) at frequency $2\omega$. Field-induced SHG has significant practical applications in visualizing electric fields.[2–6] Another example of SHG induced by symmetry breaking was recently demonstrated by some of us,[7] following a theoretical prediction,[8] where a pure spin current, i.e. a net flow of angular momenta of electrons, induces SHG.

In metals and semiconductors, the electron states are extended rather than localized. A better approach to the matter-field interaction is to use the momentum gauge with fields described by the vector potentials and the matter response by the electric currents. There is a correspondence between nonlinear phenomena in the two gauges. For instance, long-known third-order wave mixing effects in the coordinate gauge [9] have a more recently discovered analog in the momentum gauge – coherent current injection.[10, 11] Hence, it is reasonable to anticipate the existence of the field-induced SHG analog in which the symmetry is broken by the current rather than by the field. This phenomenon, current-induced SHG, was first predicted by one of us.[12] In this work, we provide conclusive experimental evidence of the current-induced SHG, and measure its strength and find it close to our theoretical estimation. We demonstrate its practical use by time-resolving a plasma oscillation and spatially resolving a steady current in GaAs samples.

We fabricate a metal-semiconductor-metal device by depositing a pair of Au electrodes on a GaAs wafer of 0.5-mm thick, as shown schematically in the inset of Fig. 1A. The electrodes are separated by a distance of about 14 $\mu$m and are approximately 1 by 2 mm in size. The wafer is n-type doped with a concentration of $10^{18}/cm^3$, grown along [100] direction. Its room-temperature resistivity $\rho = 2.3 \times 10^{-5}$ Ohm·m. Hence, a 5-V voltage across the electrodes drives a direct current of the density $J \approx 10^6$ A/cm$^2$.

In order to observe the SHG induced by this current, we use a 0.5-nJ, 170-fs, and 1800-nm probe pulse that is linearly polarized along the current direction. It is focused to a spot size of approximately 4 $\mu$m (full width at half maximum) at the same side as the electrodes by using a microscope objective lens. The transmitted SH of the probe pulse at 900 nm is collected by another objective lens, and is detected by a silicon photodiode. For lock-in detection, we modulate the current on and off by modulating the voltage applied with a square wave. In order to avoid any attenuation of the current caused by the response of the device, we use a small modulation frequency of 10 Hz. The sample is at room temperature.

In addition to the current, the sample itself also causes SHG, with a power of about 100 nW, much larger than the expected power of the current-induced SH. By comparing the surface and bulk SHG by focusing on the surface and inside, we verify that the surface contribution dominates. In our experiment, such a background is utilized as the local oscillator of the homodyne detection.[13] The total SH power is a result of the interference of these two SH fields. By modulating the current density, we can directly detect $\Delta P$, the change of the total SH power caused by the current.

We start by changing the voltage applied (and therefore the current density) and measuring the $\Delta P$ at the center of the gap between the two electrodes. Figure 1A shows that $\Delta P$ is proportional to $J$. This is consistent with the theoretical prediction that the current induced second-order susceptibility, $\chi_J^{(2)}$, is proportional to $J$.[12] Furthermore, although not shown in the figure, $\Delta P$ flips signs when the direction of the current is reversed, but retains the same amplitude.

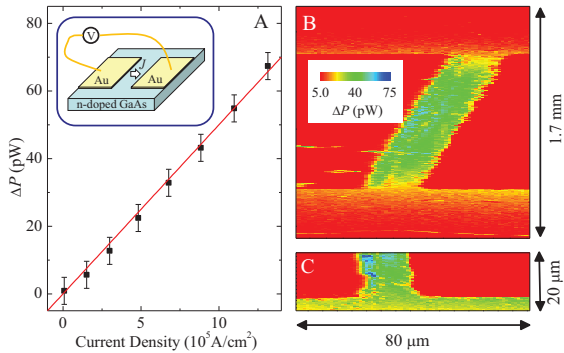

FIG. 1. SHG induced by an electrically generated steady current. A: The SH signal as a function of current density. The inset shows schematically the device geometry. B: Image of SH signal measured by scanning the probe laser spot across the device. C: Same as B, but close to the bottom edge of the electrodes.

Next, we measure $\Delta P$ at various positions throughout the device by scanning the laser spot. The result is shown in Fig. 1B. Large signals are observed in the gap between the two electrodes (the greenish strip). Figure 1C shows the same $\Delta P$ measured closer to the bottom edge of the device, and with a smaller step size. Although the geometry of this device is simple, these measurements demonstrate that the current-induced SHG can be used to obtain a real space image of current density. As such, it can be used to determine the spatial distribution of current density in a much more complex device.

In this configuration, the current is generated by the applied electric field, which is known to induce SHG.[1–6] However, in such a highly conductive sample, the field effect is expected to be small. We confirm this by a simple order-of-magnitude estimate: Our theory (Ref. 12, see also Eq. 3 and related discussions) gives the relation between nonlinear susceptibility and current density as $\chi_J^{(2)}/J \sim 2 \times 10^{-22}$ m$^3$/W under the experimental conditions. The electric field equivalent is $\chi_E^{(2)}/J = \chi^{(3)}(2\omega; 0, \omega, \omega)\rho$, where $\chi_E^{(2)}$ is the field-induced nonlinear susceptibility. The maximum plausible value of the proper third-order susceptibility $\chi^{(3)}(2\omega; 0, \omega, \omega) \sim 2.5 \times 10^{-19}$ m$^2$/V$^2$, obtained by using Miller's rule[14] and experimental data from Ref. 15. Hence, $\chi_E^{(2)}/J < 0.5 \times 10^{-23}$ m$^3$/W.

Although the field contribution is more than one order of magnitude smaller than the current contribution, an unambiguous demonstration of the current-induced SHG can only be achieved by using a current that is not constraint by Ohm's law, i.e. a current not driven by an applied field. It is possible to generate such a current by a well-know coherent current injection process, utilizing quantum interference between multiple transition pathways. In this process (Fig. 2A), a single photon of SH with frequency $2\omega$ causes transition of electrons from the valence to the conduction band. The simultaneously present two-photon transition creates electron-hole pairs too, but the most intriguing phenomenon is the interference of the two transitions that depends on the relative phase of the two fields. When they are $\pi/2$ out of phase, the interference term is positive for $+k$ and negative for $-k$. Hence more electron-hole pairs are created that move to the right as opposed to those moving to the left. The resulting non-equilibrium carrier distribution function $f_\mathbf{k}$ is different from the symmetric equilibrium function $f_\mathbf{k}^0$, as shown in Fig. 2A. Using Fermi's golden rule and performing summation over the momenta of nonequilibrium carriers, one obtains the current-injection rate,

$$\frac{dJ}{dt} = \frac{2\pi}{\hbar V}\sum_\mathbf{k}\left|\frac{eP_{\mathbf{k},x}}{m_0\omega}\right|^2 \frac{e^2\hbar^2 k_x^2}{\mu^2\hbar\omega}\mathcal{E}_\omega^2 \frac{2\mathcal{E}_{2\omega}^*}{2\omega}\delta\left(E_{cv}-2\hbar\omega\right)$$

$$= \frac{4\pi}{15}\frac{e^2 d_{cv}^2}{\mu\hbar\omega}\left(\frac{E_{cv}}{2\hbar\omega}-1\right)\rho_{cv}(2\hbar\omega)\mathcal{E}_\omega^2\mathcal{E}_{2\omega}^*, \qquad (1)$$

where $\rho_{cv}$ is the joint density of states and the dipole matrix element of the transition is defined as $d_{cv} = eP_{cv}/m_0\omega$ where $P_{cv}$ is the interband matrix element defined by Kane.[16]

To inject such a current, electrons in a 400-nm thick GaAs crystal, grown along [100], are excited from the valence band to the conduction band by one-photon absorption of a 290-fs, 750-nm pulse and two-photon absorption of a 75-fs, 1500-nm pulse. Both pulses are incident normal to the sample and are tightly focused to 2 - 3 $\mu$m at the sample surface by using a microscope objective lens. With both pulses being linearly polarized along an arbitrarily chosen $\hat{x}$ direction, electrons are excited to the conduction band with an average velocity $v_0\sin(\Delta\phi)\hat{x}$, where $\Delta\phi$ is the relative phase of the two transition amplitudes, and $v_0$ is on the order of 30 nm/ps.[11, 17–21] With a carrier density on the order of $10^{17} - 10^{18}$/cm$^3$, $J \sim 10^5$ A/cm$^2$. Since there is no driving force, the current is transient. In order to extend the lifetime of this current, the sample is cooled to 10 K.

The SHG induced by the optically injected current is observed by using an $\hat{x}$-polarized, 0.1 nJ, 170-fs, and 1760-nm probe pulse that is focused to a spot size of 2.1 $\mu$m from the backside of the sample. The SH of the probe pulse at 880 nm is collected by the pump-focusing lens, and is sent to the silicon photodiode. Similar to the DC measurement, the current-induced SH is amplified by the surface SH of 4 nW. A combination of bandpass and color filters is used in front of the photodiode in order to block the unwanted beams, including the pumps, the probe, and the photoluminescence of the sample. In addition, the photodiode is not sensitive to the strong probe at 1760 nm and the strong pump at 1500 nm.

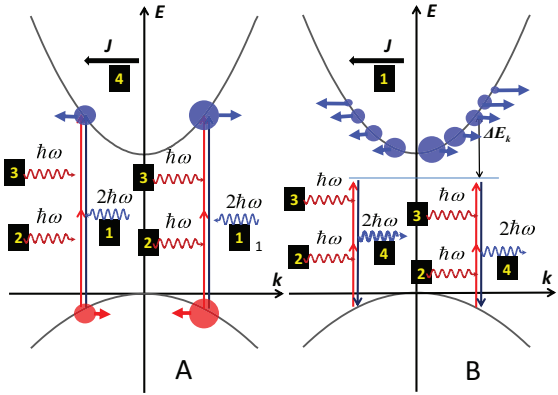

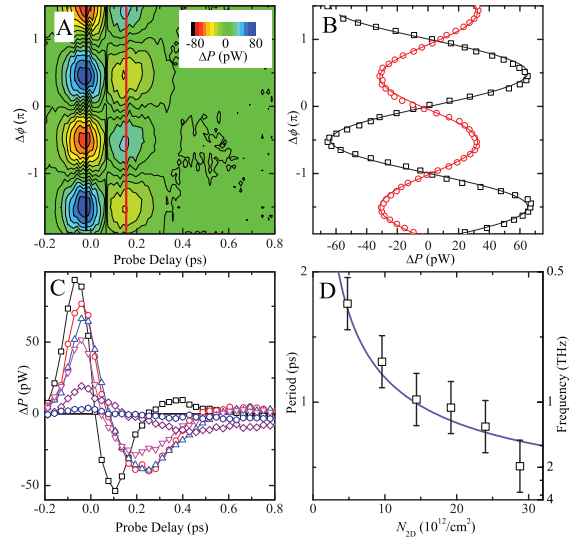

FIG. 2. A: diagrammatic representation of the sequence of events in coherent current injection. 1: single photon band-to-band transition caused by the SH photon; 2 and 3: two-photon transition caused by two fundamental photons with transition matrix elements incorporating the electron momentum; 4: interference of the two above processes results in difference of electron-hole pair generation at $+k$ and $-k$ causing current injection. B: diagrammatic representation of the sequence of events in current-induced SHG. 1: a DC current is the result of an asymmetric distribution of carriers in $k$-space; 2 and 3: virtual two-photon transition caused by two fundamental photons with transition matrix elements incorporating the electron momentum and detuning from resonance by $\Delta E_k$; 4: Polarization at two-photon frequency causes emission of a SH photon.

FIG. 3. SHG induced by an optically generated transient current. A: the measured $\Delta P$ as a function of the probe delay and $\Delta \phi$, when the probe spot overlaps with the current-injecting spots (defined as $x = 0$). The peak carrier density is $7.2 \times 10^{17}/\text{cm}^3$. B: two cross sections of Panel A with fixed probe delays of -0.02 and 0.15 ps, respectively, as indicated by the vertical lines in Panel A. C: $\Delta P$ as a function of time for several carrier densities of 7.2 (squares), 6.0 (circles), 4.8 (up-triangles), 3.6 (down-triangles), 2.4 (diamonds), and $1.2 \times 10^{17}/\text{cm}^3$ (hexagons), respectively, measured with $\Delta \phi = \pi/2$. D: the period (left axis) and the frequency (right axis) of the oscillations. The solid line indicates the $\sqrt{N_{2D}}$ dependence of the frequency.

Figure 3A shows the detected $\Delta P$ as we vary $\Delta \phi$ and the time delay between the current-injecting pulses and the probe pulse. At each probe delay, $\Delta P \propto \sin(\Delta \phi)$, as shown in Fig. 3B. Since $J \propto \sin(\Delta \phi)$, we confirm that $\chi_J^{(2)} \propto J$, which is consistent with Fig. 1A. By rotating a polarizer in front of the detector, we find that the SH is linearly polarized along $\hat{x}$ direction. Furthermore, we verify that with a $\hat{y}$-polarized probe pulse (i.e., perpendicular to the current), the $\Delta P$ is reduced by at least one order of magnitude. Hence, the SHG effect can be used to measure both the magnitude and the direction of the current density.

Figure 3A also shows that with a certain value of $\Delta \phi$, $\Delta P$ oscillates in time, as expected from a plasma oscillation: The electrons and holes are injected with opposite crystal momenta. Once they separate, a strongly nonuniform space charge field develops, which decelerates the carriers and causes the current density to drop. After the carriers reach their maximum displacements, with the current density dropping to zero, they are driven back towards the origin by the space charge field, giving rise to a negative current. As shown in Fig. 3A, such a plasma oscillation is strongly damped, due to scattering and the field inhomogeneity.[22] Furthermore, with different $\Delta \phi$, and hence different injected average velocity, the magnitude, but not the frequency, of the oscillation changes. This is consistent with the fact that the amplitude of the oscillation is determined by the initial velocity, but the frequency is independent of it.

To further investigate the plasma oscillation, we set $\Delta \phi = \pi/2$ and measure $\Delta P$ as a function of the probe delay with various peak carrier densities by varying the pump fluence, as shown in Fig. 3C. Clearly, both the magnitude and the frequency of the oscillation increase with the carrier density. Figure. 3D shows the periods and the frequencies of the oscillation. The periods are deduced by using the time difference between the first and the second zero-crossing points for each curve. Due to the large uncertainties of the data, we do not attempt to accurately analyze the dependence of the frequency on the carrier density. However, we found that the data is consistent with $\sqrt{N_{2D}}$ dependence expected for a two-dimensional plasma oscillation,[22] as indicated by the solid line. Here $N_{2D}$ is the areal carrier density. We note that although current injection by the coherent control technique has been demonstrated in many materials by steady-state electric measurement, terahertz detection, and spatially resolved pump-probe techniques,[11, 19, 20, 23–25] the current-induced SHG demonstrated here allows us to *time-resolve* the ultrafast dynamics of these currents.

Although the coherently injected current in our experiment is accompanied by a space charge field, we can safely rule out the latter as the cause of SHG: The space charge field is proportional to the charge separation, and hence is delayed with respect to $J$ by approximately a quarter period. Such a lag has been confirmed in our previous high-resolution pump-probe experiments, where the charge separation was found to reach a peak after more than 100 fs.[26, 27] However, here we observe the peak SHG around zero probe delay. Hence, the all-optical time-resolved technique has the advantage to unambiguously distinguish the field-induced and the current-induced SHG effects. The observed ultrafast dynamics also ensure that the observed signal is not merely a modification of the surface SHG by the photoexcited carriers, which would have persisted for the lifetime of the carriers of about 100 ps, and would have not shown the oscillatory features.

It is interesting to note that the current-induced SHG demonstrated above is in fact closely related to the coherent current injection process (Fig. 2A) used to inject the transient current. To illustrate this relation, we show the current-induced SHG process in Fig. 2B. Here the nonequilibrium carriers are injected first and the carrier distribution is shifted along the $k_x$ direction, with the current density $J = \sum_{c,v,\mathbf{k}} -e\hbar k_x(f_\mathbf{k} - f_\mathbf{k}^0)/m_{c,v}$. When the energy $2\hbar\omega$ is less than the bandgap, the two photon virtual upward transition is followed by a downward transition accompanied by the emission of SH photon whose rate is proportional to $J$. Normally, this SHG would not be observed since the contributions from the states with opposite $\mathbf{k}$ cancel each other. But in the presence of the current the cancelation is not complete as there are more carriers that block this process that have $-k$ as opposed to $+k$, leading to the SH polarization

$$\mathcal{P}_{2\omega}^* = -\frac{1}{2V}\sum_\mathbf{k}\left|\frac{eP_{\mathbf{k},x}}{m_0\omega}\right|^2 \frac{e\hbar k_x}{2\mu\omega\hbar\omega\Delta E_k}(f_\mathbf{k} - f_\mathbf{k}^0)\mathcal{E}_\omega^2. \quad (2)$$

One can see the unmistakable resemblance between Eqs. 1 and 2. Once summation is performed one obtains the expression for the current-induced second-order susceptibility

$$\chi_J^{(2)} = \mathcal{P}_{2\omega}^*/\varepsilon_0\mathcal{E}_\omega^2 = \frac{d_{cv}^2}{20\varepsilon_0\hbar\omega^2\Delta E_k}J. \quad (3)$$

The major difference between Eqs. 3 and 1 is that Eq. 1 describes the real process of coherent current injection with the real transition between two bands taking place – hence the presence of the density of states – while the current-induced SHG process is virtual and in place of density of states, $\Delta E_k$ – detuning averaged over all the current-carrying states contributing to the SHG – enters the expression, resulting in the Kramers-Kronig-like relation between the two processes.

To estimate from the measured $\Delta P$ the size of the nonlinearity induced by the transient current, we assume a perfect phase matching in the SHG, and solve coupled-wave equations.[14] Such a simplification is justified since the sample thickness is smaller than the coherence length. We estimate the magnitude of $\chi_J^{(2)}$ to be on the order of 0.05 pm/V with $J = 10^5 \text{A/cm}^2$. To compare with our theory, we use Eq. 3. For the experimental conditions with $d_{cv}$ evaluated using the value $2P_{cv}^2/m_0 = 28$ eV for GaAs and the average detuning taken to be $\Delta E_k \approx 0.2$ eV under the assumption of injected electrons keeping their kinetic energy, we obtain the value of $\chi_J^{(2)}/J = 7 \times 10^{-23}$ m$^3$/W. Hence, the current of $10^5$ A/cm$^2$ is expected to induce the $\chi_J^{(2)}$ of about 0.07 pm/V, which agrees very well with the experimental result.

In summary, we have demonstrated a second-order nonlinear optical effect induced by electric currents. In contrast to the recently demonstrated pure spin current-induced SHG,[7, 8] which only offers the measure of chirality of the medium, the effect demonstrated here detects electric currents, which are used in vast majority of electronic applications. The pure spin current-induced SHG relies on the existence of at least two valence bands that are split by spin-orbital interactions, while the electric current-induced SHG should exist in any type of semiconductors. Since it is an effect of the symmetry breaking at the macroscopic scale, not related to the unit cell symmetry of the crystal, it should also be observable in amorphous materials and polymers. Hence, this new member of the externally induced nonlinear optical effects, in addition to the field-induced SHG[1] and the pure spin current-induced SHG,[7, 8] can be used for direct optical detection of electric currents in a wide range of materials. Since femtosecond lasers are widely available, this ultrafast current-detection technique can be applied in many research fields to study ultrafast charge transport, as illustrated by the time resolution of the ultrafast plasma oscillation. Also, this technique can be used for a real-space image of current density, which may have applications in semiconductor industry, where such a map of the current density is required.

We thank John Prineas for providing us with high quality GaAs samples and Siyuan Han for discussions. We acknowledge financial supports from the US National Science Foundation under Awards No. DMR-0954486 and No. EPS-0903806, and matching support from the State of Kansas through Kansas Technology Enterprise Corporation, from the MCI of Spain grant FIS2009-12773-C02-01 and "Grupos Consolidados UPV/EHU del Gobierno Vasco" grant IT-472-10.